\documentclass[epj]{svjour}
\usepackage{graphics}
\begin{document}
\title{Corrections to Hawking-like Radiation for a Friedmann-Robertson-Walker Universe}
\author{Tao Zhu\thanks{Email: zhut05@lzu.cn} \and Ji-Rong Ren
 }
%
%
\institute{Institute of Theoretical Physics, Lanzhou University,
Lanzhou 730000, China}
\date{Received: date / Revised version: date}
%
\abstract{Recently, a Hamilton-Jacobi method beyond semiclassical
approximation in black hole physics was developed by \emph{Banerjee}
and \emph{Majhi}\cite{beyond0}. In this paper, we generalize their
analysis of black holes to the case of Friedmann-Robertson-Walker
(FRW) universe. It is shown that all the higher order quantum
corrections in the single particle action are proportional to the
usual semiclassical contribution. The corrections to the
Hawking-like temperature and entropy of apparent horizon for FRW
universe are also obtained. In the corrected entropy, the area law
involves logarithmic area correction together with the standard
inverse power of area term.
\PACS{~04.70.Dy} 
} 
\maketitle
Inspired by black hole thermodynamics\cite{Black Hole,ql1}, it was
realized that there is a profound connection between gravity and
thermodynamics. In \cite{Jacobson}, Jacobson first showed that the
Einstein equation can be derived from the proportionality of entropy
to the horizon area, together with the Clausius relation $\delta
Q=TdS$. Here $\delta Q$ and $T$ are the energy flux and Unruh
temperature detected by an accelerated observer just inside the
local Rindler causal horizons through spacetime point. Jacobson's
derivation has also been applied to $f(R)$ theory \cite{fr-jac} and
scalar-tensor theory\cite{sca}, where the non-equilibrium
thermodynamics must be taken into account. For other viewpoint see
\cite{other}.

With the spirit of Jacobson's derivation of Einstein field equation,
one is able to derive Friedmann equations of a FRW universe with any
spatial curvature by applying the Clausius relation to apparent
horizon of the FRW universe. For FRW universe\cite{gauss-love1},
after replacing the event horizon by the apparent horizon of FRW
universe and assuming that the apparent horizon has an associated
entropy $S_{\texttt{BH}}$ and a temperature $T_0$
\begin{eqnarray}
S_{\texttt{BH}}=\frac{A}{4\hbar},~~~T_0=\frac{\hbar}{2\pi
\tilde{r}_A},
\end{eqnarray}
one can turn the first law of thermodynamics,
$dE=T_0dS_{\texttt{BH}}$, to the Friedmann equations. Here $\hbar$,
$A$, and $\tilde{r}_A$ are the Planck constant, area of the apparent
horizon, and radius of the apparent horizon, respectively. Here it
should be noted that the entropy $S_{\texttt{BH}}$ and temperature
$T_0$ are both the semiclassical results. The first law of
thermodynamics not only holds in Einstein gravity, but also in
Gauss-Bonnet gravity, Lovelock gravity, and various braneworld
scenarios\cite{fr,gb,brane}. The fact that the first law of
thermodynamics holds extensively in various spacetime and gravity
theories suggests a deep connection between gravity and
thermodynamics. (Some other viewpoints and further developments in
this direction see \cite{zhu,padm,mass,cft/frw,brick,cao2} and
references therein.)

Since we can view a FRW thermodynamical system, like as black
holes\cite{hawking}, it is of great interest to ask that whether
there is a Hawking-like temperature associated with the apparent
horizon of FRW universe. Recently, the scalar particle and fermion's
Hawking-like radiation from apparent horizon of FRW universe were
investigated by using the semiclassical tunneling
method\cite{cao1,li}. The Hawking-like temperature $T_0=\hbar/2\pi
\tilde{r}_A$, which associated with the apparent horizon of FRW
universe, was recovered.

The semiclassical tunneling process was initially proposed by Parikh
and Wilczek\cite{Wilczek}. In recent years, it has already attracted
a lot of attention\cite{other1,other9}. In Parikh and Wilczek's
method, the imaginary part of the action is calculated with using
the null geodesic equation. In addition to the null geodesic method,
there is another method which was first developed by Padmanabhan
et.al\cite{pada5}. In this method, the Hawking radiation is derived
by calculating the particles' classical action from the
Hamilton-Jacobi equation. This method has been applied to more
general and complicated spacetimes\cite{other8} and dynamics black
holes\cite{dynamics}, and also using this method, the tunneling of a
Dirac particle through the event horizon was studied\cite{fermi}.
Later, the connection between the anomaly approach and tunneling
formulism is also discussed\cite{majhi}. Recently, the derivation of
Hawking black body spectrum in the tunneling formulism is
addressed\cite{majhi2} and this derivation fills the gap in the
existing tunneling formulations. Both the null geodesic method and
the Hamilton-Jacobi method are, however, confined to the
semiclassical approximation only. The issue of higher order quantum
corrections to the Hawking-like radiation from apparent horizon of
FRW universe is generally not discussed.

Recently, an interesting improvement has already been made by
\emph{Banerjee} and \emph{Majhi}\cite{beyond0}. They formulated the
Hamilton-Jacobi method of tunneling beyond semiclassical
approximation by considering all the terms in the expansion of the
one particle action for a scalar particle, and obtained all the
higher order quantum corrections to the semiclassical results. Some
further applications of their method to other black holes, dynamics
black holes and fermion tunneling also have been done\cite{beyond1}.
However, examples given were mostly confined to black holes.

In this paper, we generalize Hamilton-Jacobi method of tunneling
beyond semiclassical approximation of black holes to the case of FRW
universe.\footnote{We note that after submission of this manuscript,
Ref.\cite{be} appeared, which also treats the Hawking-like radiation
in FRW universe by using the tunneling method beyond semiclassical
approximation. Unlike \cite{be} only considering the corrections to
the Hawking-like temperature, we obtained the quantum corrections
both to the semiclassical Hawking-like temperature and the entropy
of apparent horizon.} We also explicitly compute all the higher
order quantum corrections to the Hawking-like temperature and the
entropy of apparent horizon of FRW universe. Let us start with the
standard FRW metric,
\begin{eqnarray}
ds^2=-dt^2+a^2\left(\frac{dr^2}{1-kr^2}+r^2d\Omega_{2}^2\right),\label{metric0}
\end{eqnarray}
where $d\Omega_{2}^2=d\theta^2+sin^2\theta d\varphi^2$ denotes the
line element of an unit two-sphere $S^2$, $a$ is the scale factor of
our universe and $k$ is the spatial curvature constant which can
take values $k=+1$ (positive curvature), $k=0$ (flat), and $k=-1$
(negative curvature). The metric (\ref{metric0}) can be rewritten as
\begin{eqnarray}
ds^2=h_{ab}dx^adx^b+\tilde{r}^2d\Omega_{2}^2,
\end{eqnarray}
where $\tilde{r}=ar$ and $x^0=t$, $x^1=r$ and the two-dimensional
metric $h_{ab}=$diag$(-1,a^2/(1-kr^2))$. In FRW universe, there is a
dynamical apparent horizon, which is the marginally trapped surface
with vanishing expansion and determined by the relation
$h^{ab}\partial_a\tilde{r}\partial_b\tilde{r}=0$. After a simple
calculation one can obtain the radius of the apparent horizon
\begin{eqnarray}
\tilde{r}_A=\frac{1}{\sqrt{H^2+k/a^2}},
\end{eqnarray}
where $H$ is the Hubble parameter, $H\equiv \dot{a}/a$ (the dot
represents derivative with respect to the cosmic time $t$). In the
tunneling approach of reference \cite{Wilczek} the
Painlev\'e-Gulstrand coordinates are used for the Schwarzschild
spacetime. Applying the change of radial coordinate, $\tilde{r}=ar$,
along with the above definitions of $H$ and $\tilde{r}_A$ to the
metric in (\ref{metric0}) one obtains the Painlev\'e-Gulstrand-like
metric for the FRW spacetime
\begin{eqnarray}
ds^2=-\frac{1-\tilde{r}^2/\tilde{r}_A^2}{1-k
\tilde{r}^2/a^2}dt^2-\frac{2 H \tilde{r}}{1-k
\tilde{r}^2/a^2}dtd\tilde{r}\nonumber\\
+\frac{1}{1-k\tilde{r}/a^2}d\tilde{r}^2+\tilde{r}^2d\Omega_2^2.\label{metric}
\end{eqnarray}
These coordinates have been used in both null geodesic method and
Hamilton-Jacobi method \cite{cao1,li} to study the Hawking-like
radiation from a FRW metric.

Consider the massless scalar field $\phi$ in the FRW universe, which
obey the Klein-Gordon equation
\begin{eqnarray}
\frac{-\hbar^2}{\sqrt{-g}}\partial_\mu(g^{\mu\nu}\sqrt{-g}\partial_\nu)\phi=0.\label{kg}
\end{eqnarray}
Since FRW universe is spherical symmetric, we only interest in the
$(t-\tilde{r})$ sector of the spacetime. By the standard ansatz for
scalar wave function
\begin{eqnarray}
\phi(\tilde{r},t)=\exp\left[\frac{i}{\hbar}S(\tilde{r},t)\right],
\end{eqnarray}
the Klein-Gordon equation (\ref{kg}) can be simplified to
\begin{eqnarray}
&&\frac{\partial^2S}{\partial
t^2}+\left(\frac{i}{\hbar}\right)\left(\frac{\partial S}{\partial
t}\right)^2+\frac{H}{1-k\tilde{r}^2/a^2}\frac{\partial S}{\partial
t}+\nonumber\\
&&\frac{\tilde{r}(H^2\tilde{r}_A^2+1-k\tilde{r}^2/a^2)}{\tilde{r}_A^2(1-k\tilde{r}^2/a^2)
}\frac{\partial S}{\partial \tilde{r}}-
\left(\frac{i}{\hbar}\right)\left(1-\frac{\tilde{r}^2}{\tilde{r}_A^2}\right)\left(\frac{\partial
S}{\partial \tilde{r}}\right)^2+\nonumber\\
&&2\frac{i}{\hbar}H\tilde{r}\frac{\partial S}{\partial
\tilde{r}}\frac{\partial S}{\partial
t}+2H\tilde{r}\frac{\partial^2S}{\partial t\partial
\tilde{r}}-\left(1-\frac{\tilde{r}^2}{\tilde{r}_A^2}\right)\frac{\partial^2S}{\partial
\tilde{r}^2}=0.\label{eq1}
\end{eqnarray}
An expression of $S(\tilde{r},t)$ in powers of $\hbar$ gives,
\begin{eqnarray}
S(\tilde{r},t)=S_0(\tilde{r},t)+\sum_i\hbar^iS_i(\tilde{r},t),\label{s}
\end{eqnarray}
where $i=1,2,3\ldots$. In the semi-classical approach, we only
consider the lowest term $S_0(\tilde{r},t)$ and neglect the terms
with $\hbar$ and greater. In this case, from (\ref{eq1}) one can get
the following equation,
\begin{eqnarray}
(\partial_t S_0)^2+2H\tilde{r}\partial_t S_0\partial_{\tilde{r}}
S_0-(1-\tilde{r}^2/\tilde{r}_A^2)(\partial_{\tilde{r}} S_0)^2=0,
\end{eqnarray}
and its solutions
\begin{eqnarray}
\partial_t S_0=(-H\tilde{r}\pm\sqrt{1-k\tilde{r}^2/a^2})\partial_{\tilde{r}}
S_0.\label{s0}
\end{eqnarray}
The higher terms with $\hbar$ and greater are treated as quantum
corrections to the semiclassical value $S_0$. Substituting (\ref{s})
into (\ref{eq1}) and using Eq.(\ref{s0}), after some calculations we
find the following relations for different powers of $\hbar$,
\begin{eqnarray}
\hbar^1: ~~~~~~~~\partial_t
S_1&=&(-H\tilde{r}\pm\sqrt{1-k\tilde{r}^2/a^2})\partial_{\tilde{r}}
S_1,\nonumber\\
\hbar^2: ~~~~~~~~\partial_t
S_2&=&(-H\tilde{r}\pm\sqrt{1-k\tilde{r}^2/a^2})\partial_{\tilde{r}}
S_2,\\
&\cdot&\nonumber\\
&\cdot&\nonumber\\
&\cdot&\nonumber
\end{eqnarray}
and so on. The above set of equations have the same functional form.
So their solutions are not independent and $S_i$ are proportional to
$S_0$. Then, we write the Eq.(\ref{s}) by
\begin{eqnarray}
S(\tilde{r},t)=(1+\sum_i\gamma_i\hbar^i)S_0(\tilde{r},t).\label{ss}
\end{eqnarray}
Here $S_0$ denotes the semiclassical contribution and the extra
value $\sum_i\gamma_i\hbar^iS_0$ can be regarded as the quantum
correction terms of the semiclassical analysis.

In order to find the solution of $S_0(\tilde{r},t)$ satisfying
Eq.(\ref{s0}), one must analysis the symmetries of the metric
(\ref{metric}). For the metric (\ref{metric}), since the metric
coefficients are both radius and time dependent, there is no time
translation Killing vector field in same with the case of static
spacetime. However, following Kodama\cite{kodama}, for spherically
symmetric dynamical spacetime whose metric like (\ref{metric}),
there is a natural analogue, the Kodama vector
\begin{eqnarray}
K=\sqrt{1-k\tilde{r}^2/a^2}\partial_t.
\end{eqnarray}
(For details of the definition of the Kodama vector and its
significance, see \cite{kodama,kodama2}.) The Kodama vector in
dynamical spacetime is of the same significance with the Killing
vector in static spacetime. It should be noted that the Kodama
vector is timelike, null and spacelike as $\tilde{r}<\tilde{r}_A$,
$\tilde{r}=\tilde{r}_A$ and $\tilde{r}>\tilde{r}_A$, respectively.
Using the Kodama vector, one can define the energy $\omega$ and
radial momentum $k_{\tilde{r}}$ measured by the Kodama observer
\begin{eqnarray}
\omega=-K\partial_tS_0=-\sqrt{1-k\tilde{r}^2/a^2}\partial_tS_0,~~~k_r=\partial_{\tilde{r}}S_0.
\end{eqnarray}
Thus one can separate $S_0$ as
\begin{eqnarray}
S_0=-\int\frac{\omega}{\sqrt{1-k\tilde{r}^2/a^2}}dt+\int
k_{\tilde{r}}d\tilde{r}.\label{antaz}
\end{eqnarray}
Substituting the above ansatz into Eq.(\ref{s0}) yields
\begin{eqnarray}
k_{\tilde{r}}=\frac{-H\tilde{r}\pm\sqrt{1-k\tilde{r}^2/a^2}}{(1-\tilde{r}^2/\tilde{r}_A^2)
\sqrt{1-k\tilde{r}^2/a^2}}\omega,\label{kr}
\end{eqnarray}
where the $+/-$ sign corresponds to the outgoing/intgoing solutions,
respectively. Therefore solutions of action $S(\tilde{r},t)$ for the
ingoing and outgoing particle under the background metric
(\ref{metric}) are respectively,
\begin{eqnarray}
S_{\texttt{out}}(\tilde{r},t)&=&\bigg[-\int\frac{\omega}{\sqrt{1-k\tilde{r}^2/a^2}}dt+\nonumber\\
&&\omega\int\frac{-H\tilde{r}+\sqrt{1-k\tilde{r}^2/a^2}}{(1-\tilde{r}^2/\tilde{r}_A^2)
\sqrt{1-k\tilde{r}^2/a^2}}d\tilde{r}\bigg]\nonumber\\
&&\times\big(1+\sum_i\gamma_i\hbar^i\big),\label{out}
\end{eqnarray}
and
\begin{eqnarray}
S_{\texttt{in}}(\tilde{r},t)&=&\bigg[-\int\frac{\omega}{\sqrt{1-k\tilde{r}^2/a^2}}dt+\nonumber\\
&&\omega\int\frac{-H\tilde{r}-\sqrt{1-k\tilde{r}^2/a^2}}{(1-\tilde{r}^2/\tilde{r}_A^2)
\sqrt{1-k\tilde{r}^2/a^2}}d\tilde{r}\bigg]\nonumber\\
&&\times\big(1+\sum_i\gamma_i\hbar^i\big).\label{in}
\end{eqnarray}

Recently, a problem in tunneling approach has been discussed in
which corresponds to a factor two ambiguity in the original Hawking
temperature\cite{2}. This ambiguity is resolved when we take into
account a temporal contribution to the imaginary part of
action\cite{time}. In Schwarzschild black hole, for the tunneling of
a particle across the event horizon the nature of the time
coordinate $t$ changes. This change indicates \cite{time} that $t$
coordinate has an imaginary part for the crossing of the horizon of
the black hole and correspondingly there will be a temporal
contribution to the imaginary part of action for the ingoing and
outgoing particles. For FRW universe, the radiation is observed by
the Kodama observer and the Kodama vector is timelike, null and
spacelike for the regions outside, on and inside the apparent
horizon, respectively. Because the energy of the particle is defined
by the Kodama vector, the discrepancy of Kodama vector inside and
outside the horizon will effect the temporal part of the action.
This means that the temporal part integral in (\ref{out}) and
(\ref{in}) also has an imaginary part. Therefore, outgoing and
ingoing probabilities are given by,
\begin{eqnarray}
P_{\texttt{out}}&=&|\phi_{\texttt{out}}|^2=\left|\exp\left[\frac{i}{\hbar}S_{\texttt{out}(\tilde{r},t)}\right]\right|^2\nonumber\\
&=&\exp\bigg[-\frac{2}{\hbar}\big(1+\sum_i\gamma_i\hbar^i\big)\bigg(-\texttt{Im}\int\frac{\omega}{\sqrt{1-kr^2}}dt\nonumber\\
&+&\omega\texttt{Im}\int
\frac{-H\tilde{r}+\sqrt{1-kr^2}}{(1-\tilde{r}^2/\tilde{r}_A^2)
\sqrt{1-kr^2}}d\tilde{r}\bigg)\bigg]\label{outp}
\end{eqnarray}
and
\begin{eqnarray}
P_{\texttt{in}}&=&|\phi_{\texttt{in}}|^2=\left|\exp\left[\frac{i}{\hbar}S_{\texttt{in}(\tilde{r},t)}\right]\right|^2\nonumber\\
&=&\exp\bigg[-\frac{2}{\hbar}\big(1+\sum_i\gamma_i\hbar^i\big)\bigg(-\texttt{Im}\int\frac{\omega}{\sqrt{1-kr^2}}dt\nonumber\\
&+&\omega\texttt{Im}\int
\frac{-H\tilde{r}-\sqrt{1-kr^2}}{(1-\tilde{r}^2/\tilde{r}_A^2)
\sqrt{1-kr^2}}d\tilde{r}\bigg)\bigg].\label{inp}
\end{eqnarray}
The contribution of the temporal part of the action to the tunneling
rate is canceled out when dividing the outgoing probability by the
ingoing probability because the temporal part is completely the same
for both the outgoing and ingoing solutions. It is no need to work
out the result of the temporal part of the action.

In the WKB approximation, the tunneling probability is related to
the imaginary part of the action as
\begin{eqnarray}
\Gamma&=&\frac{P_{\texttt{in}}}{P_{\texttt{out}}}\nonumber\\
&=&\exp\left[\frac{4\omega}{\hbar}\big(1+\sum_i\gamma_i\hbar^i\big)\texttt{Im}\int
\frac{1}{(1-\tilde{r}^2/\tilde{r}_A^2)}d\tilde{r}\right].
\end{eqnarray}
It is obvious that the integral function has a pole at the apparent
horizon. Through a contour integral, the tunneling probability of
ingoing particle now reads
\begin{eqnarray}
\Gamma=\exp\left[-\frac{2}{\hbar}\big(1+\sum_i\gamma_i\hbar^i\big)\pi
\omega\tilde{r}_A\right].
\end{eqnarray}
Now using the principle of ``detailed balance''\cite{pada5},
\begin{eqnarray}
\Gamma=\exp[-\omega/T]=\exp[-\omega/T],
\end{eqnarray}
the Hawking-like temperature associated with the apparent horizon
can be determined as
\begin{eqnarray}
T=\frac{\hbar}{2\pi
\tilde{r}_A}\big(1+\sum_i\gamma_i\hbar^i\big)^{-1}
=T_0\big(1+\sum_i\gamma_i\hbar^i\big)^{-1},\label{hawking}
\end{eqnarray}
where $T_0$ is the semiclassical Hawking-like temperature and other
terms are corrections coming from the higher order quantum effects.

In the Hawking-like temperature expression (\ref{hawking}), there
are un-determined coefficients $\gamma_i$. Since $S_0$ has the
dimension of $\hbar$, the coefficients $\gamma_i$ should have the
dimension of inverse of $\hbar^i$. In the units $G=c=k_B=1$ the
Planck constant $\hbar$ is of the order of square of the Planck
length $l_p$. Therefore, the coefficients $\gamma_i$ have the
dimension of $\tilde{r}_A^{-2}$. We can write the action $S$ as
\begin{eqnarray}
S(\tilde{r},t)=\left(1+\sum_i\frac{\alpha_i\hbar^i}{\tilde{r}_A^{2i}}\right)S_0(\tilde{r},t),
\end{eqnarray}
where $\alpha_i$ are dimensionless parameters. Now the Hawking-like
temperature (\ref{hawking}) can be written as
\begin{eqnarray}
T=T_0\left(1+\sum_i\frac{\alpha_i\hbar^i}{\tilde{r}_A^{2i}}\right)^{-1}.\label{hawking2}
\end{eqnarray}
Till now, we have obtained all the corrections to the semiclassical
Hawking-like temperature $T_0$. It should be noted that the Kodama
observer is inside the apparent horizon. This means that the Kodama
observer does see a thermal spectrum with temperature $T=(1/2\pi
\tilde{r}_A)(1+\sum_i\frac{\alpha_i\hbar^i}{\tilde{r}_A^{2i}})^{-1}$.

Hawking temperature is always related to surface gravity of horizon
as $T=\kappa/2\pi$. In the semiclassical case, the surface gravity
is $\kappa_0=1/\tilde{r}_A$. Hence, the modified form of surface
gravity of the apparent horizon following from (\ref{hawking2}), is
\begin{eqnarray}
\kappa=\kappa_0\left(1+\sum_i\frac{\alpha_i\hbar^i}{\tilde{r}_A^{2i}}\right)^{-1}.
\end{eqnarray}

Now let us turn to investigate the entropy of apparent horizon in
the presence of higher order quantum corrections. The semiclassical
Bekenstein-Hawking entropy of apparent horizon is given by
\begin{eqnarray}
S_{\texttt{BH}}=\frac{A}{4\hbar}.\label{entropy}
\end{eqnarray}
The first law of thermodynamics holds on apparent horizon indicates
$dS_{\texttt{BH}}=dE/T_0$, where $dE$ is the amount of energy
crossing the apparent horizon in FRW universe. In constructing the
first law of thermodynamics on apparent horizon, a key point is to
calculate this energy $dE$ in an infinitesimal time interval. In FRW
universe, the total energy inside the apparent horizon is defined by
a quasi-local mass: the Misner-Sharp mass $M=\tilde{r}_A/2$. By
using the Misner-Sharp mass $M$, the energy flux passed through the
apparent horizon is defined as
\begin{eqnarray}
dE=(k^t\partial_tM+k^r\partial_rM)dt=d\tilde{r}_A,
\end{eqnarray}
where $k^{t,r}=(1,-Hr)$ is the (approximate) generator of the
apparent horizon and satisfies
$k^r\partial_r\tilde{r}+k^t\partial_t\tilde{r}=0$. With the
expression of modified Hawking-like temperature (\ref{hawking2}),
the first law of thermodynamics on apparent horizon is
\begin{eqnarray}
dS=\frac{dE}{T}=\frac{d\tilde{r}_A}{T_0}\left(1+\sum_i\frac{\alpha_i\hbar^i}{\tilde{r}_A^{2i}}\right).
\end{eqnarray}
Integrating the above equation yields the entropy of apparent
horizon
\begin{eqnarray}
S&=&\int\frac{d\tilde{r}_A}{T_0}\left(1+\sum_{i=1}\frac{\alpha_i\hbar^i}{\tilde{r}_A^{2i}}\right)\nonumber\\
&=&\frac{A}{4\hbar}+\pi \alpha_1\ln
\frac{A}{4\hbar}+\sum_{i=2}\frac{\pi^i\alpha_i}{1-i}(\frac{A}{4\hbar})^{1-i}+\texttt{const}.\label{entropy1}
\end{eqnarray}
We can see that the first term is the usual semiclassical area law
(\ref{entropy}), and the other terms are the quantum corrections.
For the correction term, it contain two parts: the logarithmic term
and the inverse area terms. We note that the logarithmic correction
term has been also obtained by other approaches\cite{zhu,log1,log2},
and in some literatures\cite{log2}, the coefficient of the
logarithmic correction term is controversial. In our result, the
coefficient of the logarithmic correction term is determined by the
dimensionless constant $\alpha_1$.

In conclusion, we generalize Hamilton-Jacobi method of tunneling
beyond semiclassical approximation of black holes to the case of FRW
universe. We have considered all orders in the single particle
action for particle tunneling through the apparent horizon of the
FRW universe. It is shown that higher order correction terms of the
action are proportional to the semiclassical contribution. By
applying the dimensional argument and principle of ''detailed
balance'', higher order corrections to the Hawking-like temperature
and entropy of apparent horizon are obtained. For the corrected
entropy (\ref{entropy1}), it contains three parts: the usual
Bekenstein-Hawking entropy, the logarithmic term and the inverse
area term. We find that the coefficient of the logarithmic
correction term is determined by the dimensionless constant
$\alpha_1$.

\section*{Acknowledgements}

 Thank Dr. \emph{Ming-Fan Li} for his works in
improving the English Writing of this manuscript. This work was
supported by the National Natural Science Foundation of China (No.
10275030) and Cuiying Project of Lanzhou University(No.
225000-582404). We thank Dr.Elias Vagenas for pointing out a defect
of our paper.

\end{document}